\documentclass[conference,letterpaper]{IEEEtran}
\renewcommand{\IEEEkeywordsname}{Keywords}

\makeatletter
\@ifundefined{mycopyrightnotice}
{\newcommand{\mycopyrightnotice}{}}
{\renewcommand{\mycopyrightnotice}{}}
\makeatother

\usepackage[utf8]{inputenc}
\usepackage[T1]{fontenc}
\usepackage{eso-pic}
\IEEEoverridecommandlockouts
\usepackage{cite}
\usepackage{amsmath,amssymb,amsfonts}
\usepackage{graphicx}
\usepackage{tabularx}
\usepackage{array}
\usepackage{booktabs}
\usepackage{textcomp}
\usepackage{xcolor}
\usepackage{xurl}
\usepackage{placeins}
\usepackage{float}

\newcommand{\ACDSAcopyright}{979-8-3195-2824-7/27/\$31.00~\copyright~2027 IEEE}
\makeatletter
\def\ps@IEEEtitlepagestyle{\def\@oddfoot{\mycopyrightnotice}\def\@evenfoot{}}
\def\mycopyrightnotice{{\footnotesize\ACDSAcopyright\hfill}\gdef\mycopyrightnotice{}}
\makeatother
\newcommand\AtPageUpperMyright[1]{\AtPageUpperLeft{%
  \put(\LenToUnit{0.17\paperwidth},\LenToUnit{-2cm}){%
    \parbox{0.9\textwidth}{\raggedleft\fontsize{8}{11}\selectfont #1}}}}
\newcommand{\conf}[1]{\AddToShipoutPictureBG*{\AtPageUpperMyright{#1}}}
\newcommand{\tablestyle}{\footnotesize\setlength{\tabcolsep}{3.5pt}\renewcommand{\arraystretch}{1.12}}
\newcolumntype{L}{>{\raggedright\arraybackslash}X}
\begin{document}

\title{\vspace*{1cm}Tracking and Limiting Unsupported Claim Propagation in Multi-Agent LLM Systems: A Provenance-Aware Framework}

\author{
\IEEEauthorblockN{Zijia Cao}
\IEEEauthorblockA{
The University of Chicago\\
Chicago, IL, USA\\
zijcao@uchicago.edu
}
\and
\IEEEauthorblockN{Hanqing Dou}
\IEEEauthorblockA{
University of California San Diego\\
San Diego, CA, USA\\
hdou@ucsd.edu
}
}
\IEEEaftertitletext{\vspace{-\baselineskip}}
\maketitle

\conf{\textit{Proc. of International Conference on Artificial Intelligence, Computer, Data Sciences and Applications (ACDSA 2027)\\2--4 February 2027, Rio de Janeiro-Brazil}}

\begin{abstract}
This research examines how unsupported claims change as they move through multi-agent pipelines. It also examines whether preserving provenance can reduce these changes. We test both homogeneous and heterogeneous pipelines. Homogeneous pipelines use the same model across agents, while heterogeneous pipelines use different models. We compare the baseline with a provenance-aware framework. The framework passes each claim with its support status, source, evidence and transmission history. No new evidence is introduced during transmission. We measure changes in claim meaning, unsupported detail accretion, semantic drift, and provenance preservation. These metrics show whether the framework limits unsupported changes as claims pass between agents.
\end{abstract}


\section{Introduction}
Large Language Models (LLMs) are good at answering questions and generating information, but they can also return claims that are not supported by reliable evidence sometimes. It becomes difficult to determine if the claim is trustworthy when its source is not provided. This issue becomes more serious in multi-agent systems, where information is repeatedly passed between agents. An unsupported claim might be rephrased, expanded or altered during each transmission. It then becomes more difficult to tell if a piece of information comes from an earlier agent or is introduced by a later agent. This lack of traceability makes it harder for agents and users to identify and validate unsupported claims. Our research examines how unsupported claims change across five sequential agent stages, how these changes differ between homogeneous pipelines and a fixed heterogeneous pipeline and how reliably agents preserve the provenance information provided to them. We also compare provenance-aware transmission with text-only transmission to see whether the framework reduces meaning changes and unsupported additions. 

\section{Background and Current Limitation}
\subsection{LLM Hallucination}

As LLMs become popular and usage increases, researchers have paid more attention to hallucinations in model outputs. Hallucination occurs when an LLM generates content without reliable supporting evidence. Research has shown that factual reliability remains a challenge for LLMs. In TruthfulQA, the best model answered 58\% of questions truthfully while humans achieved 94\%. The study also finds that larger models do not necessarily provide more truthful answers \cite{ref1}. Researchers have studied hallucination from different perspectives, including hallucination detection, factual evaluation and evidence verification.

Hallucination detection methods differ in what they compare and whether they use external evidence. SelfCheckGPT generates several responses from a black-box model and compares them without relying on an external database. Facts known by the model are expected to remain fairly consistent, while hallucinated details are more likely to vary or conflict \cite{ref2}. FActScore instead divides long-form responses into atomic facts and checks how many are supported by a reliable source. ChatGPT scored 58\% in its evaluation of biographical generations \cite{ref3}. Both methods focus on outputs generated by an individual model. Our study examines what happens after an unsupported claim becomes another agent\textquoteright{}s input and continues through an agent pipeline.

\subsection{Multi-Agent Factual Reliability}

Researchers have studied factual reliability in multi-agent systems from different perspectives. Jamshidi et al. track how claims change as they pass between agents \cite{ref4}. They find that hallucination scores can decrease while factual accuracy also declines slightly. This means that a lower hallucination score does not always indicate a more accurate claim. Zhou et al. focus on how agents assess information received from other agents \cite{ref5}. Their Epistemic Context Learning method uses previous interactions to build reliability profiles for agents. Rashkin et al. focus on source support \cite{ref6}. Their AIS framework evaluates whether an identified source supports a generated claim. Overall, these studies examine three related aspects of factual reliability: changes in claims, trust between agents and support from external sources.

Related work in predictive systems examines how information is represented and used in downstream decisions. Chang et al. combine temporal modeling with Bayesian uncertainty estimation for credit risk prediction under incomplete observations \cite{ref7}. Yan incorporates temporal information and relational tokens into transaction representations for fraud detection \cite{ref8}. Zhang et al. separate fraud risk ranking from cost sensitive case selection under review capacity constraints \cite{ref9}. These studies provide a broader context for examining how available information shapes predictions and subsequent decisions. 

Our research focuses on the information passed between agents in a sequential multi-agent pipeline. Based on the earlier discussion of claim transformation, agent reliability and source support, we track claim content and provenance separately to examine both how a claim changes and whether its support status, source, evidence and transmission history are retained during transmission. We compare the provenance-aware framework with text-only transmission to examine whether it reduces changes in meaning and unsupported additions. 

\section{Experiment Design}

\subsection{Research Objective}

This experiment consists of two parts. First, it examines how unsupported claims change as they go through multiple agents. It checks whether the original claim is preserved and how it changes during transmission. This behavior is tested in homogeneous scenarios, which use the same model at every stage. Heterogeneous scenarios use different models to determine whether the pattern is consistent across model configurations. Second, we evaluate whether the proposed provenance-aware framework reduces unsupported claim propagation by preserving each claim\textquoteright{}s support status, source, evidence and transmission history. No new evidence is introduced at any stage.

\subsection{Dataset}

FEVER (Fact Extraction and VERification) is a fact checking dataset containing 185,445 claims evaluated against information from Wikipedia \cite{ref10}. The experiment uses a fixed set of claims labeled ``NOT ENOUGH INFO'' in FEVER because the available evidence neither supports nor contradicts them. Every test uses the same ordered dataset as the input to ensure consistency and allow direct comparison across different experimental conditions.

\subsection{Model Selection}

The experiment uses five open-weight models: Gemma 3 4B, Granite 3.3 8B, Llama 3.1 8B, Mistral 7B Instruct, and Qwen3 8B. We record the checkpoint, version, quantization format and artifact identifier for each model to support reproducibility. Local execution provides consistent control over temperature, top-p, context length, output length, and random seeds. Models from different developers allow us to examine whether unsupported claim propagation occurs across models.

\subsection{Experiment Conditions}

The experiment passes each claim through five stages, with one agent at each stage. An agent is a separate LLM inference call that processes the input and produces a summary for the next stage. Each call starts with a fresh message context and has no persistent conversational memory. 

The experiment includes four scenarios based on model configuration and the information available to each agent: homogeneous baseline, heterogeneous baseline, homogeneous provenance-aware, and heterogeneous provenance-aware. In the homogeneous scenarios, agents use the same model at all five stages. We run this pipeline separately for each of the five models. For example, one homogeneous pipeline uses Qwen at all five stages: Qwen $\rightarrow$ Qwen $\rightarrow$ Qwen $\rightarrow$ Qwen $\rightarrow$ Qwen. In the heterogeneous scenarios, the pipeline uses five different models in a fixed order: Qwen $\rightarrow$ Gemma $\rightarrow$ Llama $\rightarrow$ Mistral $\rightarrow$ Granite. Every agent is given the same task: generate a concise verified summary for the next agent.  Although the task uses the word ``verified,'' the experiment does not provide new evidence or independently verify the factual truth of each summary.  In the baseline scenarios, each agent receives only the previous agent\textquoteright{}s text and returns a plain-text summary. In the provenance-aware scenarios, each agent receives a structured prompt including the claim, its support status, source, evidence and transmission history. The agent returns the claim text and provenance fields as a JSON object, while the framework records how the claim passes through the five stages. 

All four scenarios use the same claims, a five-stage agent pipeline and generation settings: a temperature of 0.7, top-p of 0.9, a maximum output length of 256 tokens and a context length of 4,096 tokens. The baseline and provenance-aware scenarios use the same models in the same order. The experiment processes each claim three times using random seeds of 0, 26 and 42 to capture variation while keeping the results reproducible. For each claim and seed, the baseline and provenance-aware scenarios each generate five homogeneous runs and one heterogeneous run.

\subsection{Evaluation Metrics}

\label{sec:evaluation-metrics}

Let $C_{0}$ represent the original unsupported claim and $ C_{i}$ represent the claim after agent $ i$. The evaluation compares each stage\textquoteright{}s output with the original claim to measure whether the original proposition is retained, how its meaning changes and whether unsupported details are added. These comparisons measure change relative to the original claim rather than change only from the preceding stage. Retaining the original proposition does not mean that it is factually supported and contradicting an unsupported claim does not necessarily mean that the output is false. 

\subsubsection{Five-Class Bidirectional NLI Relation Evaluator}

When a claim passes from one agent to the next, we need to know not only whether it changed, but also how it changed. Previous studies show that Natural Language Inference (NLI) can be used for this purpose. Du\v{s}ek and Kasner test entailment in both directions to distinguish unsupported additions from missing information \cite{ref11}. Laban et al. instead combine sentence-level NLI results into an overall consistency score for a document \cite{ref12}. In our evaluation, we keep the two directions separate so that preservation, expansion, contraction, contradiction and semantic loss can be reported individually.

We use the pretrained MoritzLaurer/DeBERTa-v3-base-mnli-fever-anli model \cite{ref13}. For each pair of claims, the model returns probabilities for entailment, neutrality and contradiction. We compare each agent output $C_{i}$ with the original claim $ C_{0}$ in both directions. The probability $ p_{ent}(C_{i},C_{0})$indicates whether the output preserves the original claim, while $ p_{ent}(C_{0},C_{i})$ measures whether the original claim includes the information in the output. We also use $ p_{con}(C_{i},C_{0})$ to measure whether the output contradicts the original claim. The classification uses two decision thresholds: $ t_{E}$ for entailment and $ t_{C}$ for contradiction. Their calibrated values are explained after the classification rules.

\begin{table*}[t]
\caption{Five-Class Bidirectional NLI Relation Rules}
\label{tab:nli-rules}
\centering
\tablestyle
\begin{tabularx}{\textwidth}{@{}p{0.12\textwidth}p{0.34\textwidth}X@{}}
\toprule
Category & Decision Rule & Interpretation \\
\midrule
Contradicted & $p_{con}(C_{i},C_{0}) \geq$ $t_{C}$  & The agent output conflicts with the original claim. \\
Equivalent & $p_{ent}(C_{i},C_{0})\geq t_{E}$ and $p_{ent}(C_{0},C_{i})\geq t_{E}$ & The output and original claim express substantially the same proposition. \\
Expanded & $p_{ent}(C_{i},C_{0})\geq t_{E}$ and $p_{ent}(C_{0},C_{i}) <t_{E}$ & The output preserves the original proposition but adds information. \\
Contracted & $p_{ent}(C_{i},C_{0})<t_{E}$ and $p_{ent}(C_{0},C_{i})\geq t_{E}$ & The output retains only part of the original proposition. \\
Drifted/Lost & $p_{ent}(C_{i},C_{0})<t_{E}$ and $p_{ent}(C_{0},C_{i})<t_{E}$ & The original proposition is neither preserved nor directly contradicted.  \\
\bottomrule
\end{tabularx}
\end{table*}

The categories are assigned in two steps. It first checks contradiction so that an output that rejects the original claim is not classified as an expansion or contraction. Although contradiction is logically symmetric, NLI scores can change when the premise and hypothesis are reversed. During calibration, allowing contradiction in either direction to determine the label generated additional false contradictions. We therefore use $p_{con}(C_{i},C_{0})$ as the controlling contradiction score. The four entailment-based categories are considered only when $p_{con}(C_{i},C_{0}) < t_{C}$. Table~\ref{tab:nli-rules} summarizes the five-class bidirectional NLI decision rules.

We calibrate the NLI thresholds using 145 agent outputs and pair each output with its comparison claim. An LLM judge first assigned one of the five relation labels and a human reviewer checks and corrects the labels. We then test different entailment and contradiction thresholds and compare the resulting classifications with the reviewed labels. As a result, the combination $t_{E}$=0.30 and $ t_{C}$=0.65 achieved the highest accuracy.

Each claim comparison is assigned to one of the five relation categories. We then calculate the percentage of claims falling into each category for every experimental condition, model and stage. These distributions show how often claim content is preserved, expanded, reduced, contradicted, or lost during transmission.

\subsubsection{Unsupported Detail Accretion (UDA)}

UDA measures how much new unsupported information an agent adds. We use $C_{i}$ to represent a claim after agent $i$. For example, $ C_{0}$ is the original claim and $ C_{5}$ is the version generated after the fifth agent.

\begin{equation}\resizebox{0.88\columnwidth}{!}{$\displaystyle UDA_{i}=\frac{\text{number of unsupported factual units in }C_{i}}{\max(\text{number of factual units in }C_{i},1)}$}\end{equation}

In this formula, UDA is the share of the output that includes unsupported added details. The $\max(\text{number of factual units in }C_{i},1)$ term prevents division by zero when an output has no identifiable factual unit.

We split each output into smaller pieces of information using sentence boundaries, semicolons, line breaks and coordinated clauses. Each piece is treated as a candidate factual unit because it may contain a separate claim about an event, property, relationship, or source. For example, ``The film was released in 2002 and won three awards'' contains two units: one about its release date and one about its awards.

We then use the NLI evaluator to check each unit against the original claim and any available evidence. A source is treated as supporting a unit when the NLI entailment probability is at least 0.50. A unit is counted as unsupported only when neither the original claim nor the available evidence meets this threshold. UDA therefore does not decide whether an entire output is true or false. Instead, it measures how much unsupported information is added as a claim moves through the agent pipeline. A higher UDA indicates more unsupported additions.

\subsubsection{Semantic Drift}

Semantic Drift measures the overall semantic distance between the original claim and an agent output. The evaluator converts $C_{0}$ and $ C_{i}$ into sentence embeddings using the pretrained all-MiniLM-L6-v2 \cite{ref14}, \cite{ref15}. It then calculates cosine similarity between the two embeddings and defines Semantic Drift as:

\begin{equation}SD_{i}=1-\cos(e(C_{0}),e(C_{i}))\end{equation}

$e(C_{i})$ represents the sentence embedding of claim after agent $ i$. A lower SD indicates that the output remains closer in overall meaning to the original claim, while a higher SD indicates greater semantic change. We report SD as a supporting metric rather than the primary measure of claim transformation. It provides one continuous measure of overall semantic distance.

\subsubsection{Provenance Preservation Rate (PPR)}

For provenance-aware conditions, Provenance Preservation Rate measures whether each agent\textquoteright{}s output correctly retains the provenance information supplied at that stage. An output passes only when it satisfies all five checks described below. PPR is calculated separately for each stage and does not measure whether provenance was preserved in every preceding output along the same chain. 

\begin{equation}\resizebox{0.88\columnwidth}{!}{$\displaystyle PPR_{i}=\frac{\#\{\text{claims with complete provenance at stage }i\}}{\#\{\text{claims evaluated at stage }i\}}$}\end{equation}

An output is counted as having complete provenance only when five checks pass. First, the agent\textquoteright{}s response must contain valid JSON with a non-empty claim-text field. Second, the reported support status must match the status received by the agent. Third, the agent must report the same source it received. Fourth, it must preserve the supplied evidence. Finally, the lineage entries and transmission count must match the current agent stage. Source and evidence checks ignore differences in capitalization, spacing and evidence order. The metadata given to the agent provides the expected values for these checks, the process does not make a new judgment about whether the claim is true. An output has complete provenance only if it passes all five checks.

\section{Unsupported Claim Propagation in Baseline}

The baseline tracks how unsupported claims change as agents pass them from one stage to the next without provenance information. We use these results as a reference when evaluating the provenance-aware framework. The comparison is based on the metrics defined in Section~\ref{sec:evaluation-metrics}, which measure how the claim changes during transmission.

\subsection{Homogeneous Baseline}

In the homogeneous baseline, the same model processes a claim at all five stages. We repeat this setup for each of the five models and then track changes in meaning and the occurrence of unsupported details. Table~\ref{tab:homogeneous-baseline} reports the homogeneous baseline results by pipeline stage.

The homogeneous baseline shows that claim content changes as information moves through the five-stage pipeline. The percentage of outputs labeled Equivalent falls from 74.73\% at stage one to 62.67\% at stage five. At the same time, Expanded increases from 9.99\% to 14.67\%, and Drifted/Lost increases from 2.27\% to 7.43\%. Contracted and Contradicted show smaller changes. Semantic Drift increases from 0.062 to 0.104, while UDA increases from 0.207 to 0.311. These results show that claim changes and unsupported details accumulate across stages even when the agents use the same model.

\begin{table}[H]
\caption{Homogeneous Baseline Results by Pipeline Stage}
\label{tab:homogeneous-baseline}
\centering
\scriptsize
\setlength{\tabcolsep}{2.0pt}
\renewcommand{\arraystretch}{1.08}
\begin{tabular}{@{}lrrrrr@{}}
\toprule
Metric & \shortstack{Stage1} & \shortstack{Stage2} & \shortstack{Stage3} & \shortstack{Stage4} & \shortstack{Stage5} \\
\midrule
Equivalent (\%) & 74.73 & 68.91 & 65.84 & 63.99 & 62.67 \\
Expanded (\%) & 9.99 & 12.92 & 14.20 & 14.53 & 14.67 \\
Contracted (\%) & 3.03 & 3.39 & 3.56 & 3.56 & 3.87 \\
Contradicted (\%) & 9.99 & 10.71 & 11.12 & 11.45 & 11.37 \\
Drifted/lost (\%) & 2.27 & 4.08 & 5.28 & 6.47 & 7.43 \\
Semantic Drift & 0.062 & 0.080 & 0.091 & 0.098 & 0.104 \\
UDA & 0.207 & 0.258 & 0.285 & 0.301 & 0.311 \\
\bottomrule
\end{tabular}
\end{table}

\subsection{Heterogeneous Baseline}

In the heterogeneous baseline, each claim passes through five models in a fixed order ( Qwen $\rightarrow$ Gemma $\rightarrow$ Llama $\rightarrow$ Mistral $\rightarrow$ Granite). Each model receives the output from the previous model. This allows us to track how unsupported claims change as they pass across different models. Table~\ref{tab:heterogeneous-baseline} reports the heterogeneous baseline results by pipeline stage.

The percentage of outputs labeled Equivalent falls from 93.53\% at stage one to 31.33\% at stage five. Expanded increases from 3.47\% to 28.93\% while Contradicted increases from 1.93\% to 20.67\%. Drifted/Lost also increases from 0.33\% to 12.33\%. Semantic Drift increases from 0.019 to 0.194, and UDA increases from 0.060 to 0.590. These results show that unsupported claims change substantially as they pass through different models, with less than one third remaining Equivalent at the final stage.

\begin{table}[H]
\caption{Heterogeneous Baseline Results by Pipeline Stage}
\label{tab:heterogeneous-baseline}
\centering
\scriptsize
\setlength{\tabcolsep}{2.0pt}
\renewcommand{\arraystretch}{1.08}
\begin{tabular}{@{}lrrrrr@{}}
\toprule
Metric & \shortstack{Stage1\\Qwen} & \shortstack{Stage2\\Gemma} & \shortstack{Stage3\\Llama} & \shortstack{Stage4\\Mistral} & \shortstack{Stage5\\Granite} \\
\midrule
Equivalent (\%) & 93.53 & 90.33 & 79.27 & 41.60 & 31.33 \\
Expanded (\%) & 3.47 & 3.80 & 3.53 & 22.80 & 28.93 \\
Contracted (\%) & 0.73 & 2.47 & 5.33 & 7.80 & 6.73 \\
Contradicted (\%) & 1.93 & 2.07 & 8.87 & 19.73 & 20.67 \\
Drifted/lost (\%) & 0.33 & 1.33 & 3.00 & 8.07 & 12.33 \\
Semantic Drift & 0.019 & 0.028 & 0.057 & 0.143 & 0.194 \\
UDA & 0.060 & 0.071 & 0.148 & 0.466 & 0.590 \\
\bottomrule
\end{tabular}
\end{table}

\subsection{Summary}

The baseline results show that the proportion of outputs equivalent to the original claim decreases across the five stages, while semantic drift and unsupported detail accretion increase. Some outputs preserve or expand the original proposition, while others contract, contradict or lose it. These changes are larger in the tested heterogeneous sequence than in the aggregated homogeneous results. However, the heterogeneous sequence uses a fixed model order, so the results do not separate model differences from stage position and preceding transformations.  The provenance-aware framework aims to reduce these changes by keeping support information attached to each claim as it passes from one agent to the next.

\section{Provenance-Aware Framework}

In the provenance-aware framework, each agent receives a structured record containing the claim\textquoteright{}s origin, support status, source, evidence and transmission history. We compare the framework with the baseline to see whether provenance reduces changes in the claim and preserves the provenance attached to each claim.

\subsection{Homogeneous Provenance-Aware Framework}

In the homogeneous provenance-aware setting, the same model processes each claim at all five stages and every agent receives the claim\textquoteright{}s provenance information. We repeat this setup for each of the five models and examine whether the claim retains its meaning and unsupported status. We also measure whether fewer unsupported details are added as the claim moves through the chain. The results are compared with the corresponding homogeneous baseline for the same model. Table~\ref{tab:homogeneous-provenance} reports the homogeneous provenance-aware results by pipeline stage.

The homogeneous provenance-aware results remain stable across all five stages. More than 96\% of outputs remain equivalent to the original claim at every stage. Expanded outputs remain below 0.6\% and contracted outputs remain at or below 0.04\%. Semantic drift stays at or below 0.006, while unsupported detail accretion remains at or below 0.021. Contradicted and drifted or lost outputs also remain below 1.6\% and 1\%, respectively. Provenance preservation ranges from 84.32\% to 93.57\%. These results show that the framework helps agents preserve the claim\textquoteright{}s meaning and provenance while limiting unsupported changes.

\begin{table}[H]
\caption{Homogeneous Provenance-Aware Results by Pipeline Stage}
\label{tab:homogeneous-provenance}
\centering
\scriptsize
\setlength{\tabcolsep}{2.0pt}
\renewcommand{\arraystretch}{1.08}
\begin{tabular}{@{}lrrrrr@{}}
\toprule
Metric & \shortstack{Stage1} & \shortstack{Stage2} & \shortstack{Stage3} & \shortstack{Stage4} & \shortstack{Stage5} \\
\midrule
Equivalent (\%) & 98.07 & 97.25 & 97.13 & 96.97 & 96.95 \\
Expanded (\%) & 0.31 & 0.41 & 0.48 & 0.51 & 0.49 \\
Contracted (\%) & 0.01 & 0.01 & 0.03 & 0.03 & 0.04 \\
Contradicted (\%) & 0.99 & 1.43 & 1.45 & 1.56 & 1.57 \\
Drifted/Lost (\%) & 0.63 & 0.89 & 0.91 & 0.93 & 0.95 \\
Semantic Drift & 0.003 & 0.005 & 0.005 & 0.005 & 0.006 \\
UDA & 0.014 & 0.019 & 0.020 & 0.021 & 0.021 \\
\shortstack[l]{Provenance Preserved} (\%) & 90.51 & 84.32 & 91.92 & 93.57 & 91.45 \\
\bottomrule
\end{tabular}
\end{table}

\subsection{Heterogeneous Provenance-Aware Framework}

This scenario uses the same fixed sequence of different models, but every agent also receives the claim\textquoteright{}s provenance information. It tests whether the models preserve this information and whether the framework reduces unsupported changes as claims pass between models. The results are compared with the heterogeneous baseline.Table~\ref{tab:heterogeneous-provenance} reports the heterogeneous provenance-aware results by pipeline stage.

During the first three stages of the heterogeneous provenance-aware setting, every output remains equivalent to the original claim. Provenance preservation is 100\% for Qwen and Gemma and 99.67\% for Llama. At stage four, Mistral preserves the original meaning in 95.80\% of outputs and the complete provenance information in 64.60\%. At stage five, 95.73\% of Granite outputs receive the Equivalent label while 99.93\% pass all provenance checks. The higher stage-five PPR indicates better compliance with the metadata supplied at that stage, it does not indicate uninterrupted preservation across all earlier stages. At stage five, semantic drift is 0.007 and UDA is 0.029.  These results show that the framework helps preserve claims as they pass between different models. The stage-four Mistral outputs have the lowest PPR, but this result does not separate model identity from pipeline position and preceding transformations. 

\begin{table}[H]
\caption{Heterogeneous Provenance-Aware Results by Pipeline Stage}
\label{tab:heterogeneous-provenance}
\centering
\scriptsize
\setlength{\tabcolsep}{2.0pt}
\renewcommand{\arraystretch}{1.08}
\begin{tabular}{@{}lrrrrr@{}}
\toprule
Metric & \shortstack{Stage1\\Qwen} & \shortstack{Stage2\\Gemma} & \shortstack{Stage3\\Llama} & \shortstack{Stage4\\Mistral} & \shortstack{Stage5\\Granite} \\
\midrule
Equivalent (\%) & 100.00 & 100.00 & 100.00 & 95.80 & 95.73 \\
Expanded (\%) & 0.00 & 0.00 & 0.00 & 0.87 & 0.67 \\
Contracted (\%) & 0.00 & 0.00 & 0.00 & 0.13 & 0.13 \\
Contradicted (\%) & 0.00 & 0.00 & 0.00 & 1.80 & 2.07 \\
Drifted/Lost (\%) & 0.00 & 0.00 & 0.00 & 1.40 & 1.40 \\
Semantic Drift & 0.000 & 0.000 & 0.000 & 0.007 & 0.007 \\
UDA & 0.000 & 0.000 & 0.000 & 0.027 & 0.029 \\
\shortstack[l]{Provenance Preserved} (\%) & 100.00 & 100.00 & 99.67 & 64.60 & 99.93 \\
\bottomrule
\end{tabular}
\end{table}

\section{Discussion}

\subsection{Homogeneous Comparison}

Table~\ref{tab:homogeneous-comparison} compares the homogeneous baseline and provenance-aware results at stage five. At stage five, 96.95\% of provenance-aware outputs remain equivalent to the original claim, compared with 62.67\% in the homogeneous baseline. Expanded outputs decrease from 14.67\% to 0.49\%, contracted outputs from 3.87\% to 0.04\%, contradictions from 11.37\% to 1.57\% and drifted or lost outputs from 7.43\% to 0.95\%. The framework also reduces semantic drift by 94.7\% and unsupported detail accretion by 93.2\%, while preserving provenance in 91.45\% of outputs. These differences show that provenance helps preserve the original meaning and reduces cumulative unsupported changes when the same model repeatedly processes a claim.

\begin{table}[H]
\caption{Homogeneous Baseline and Provenance-Aware Comparison at Stage Five}
\label{tab:homogeneous-comparison}
\centering
\tablestyle
\begin{tabularx}{\columnwidth}{@{}Lrrr@{}}
\toprule
Metrics & Baseline & \shortstack{Provenance-\\Aware} & Difference \\
\midrule
Equivalent & 62.67\% & 96.95\% & +34.28\% \\
Expanded & 14.67\% & 0.49\% & -14.18\% \\
Contracted & 3.87\% & 0.04\% & -3.83\% \\
Contradicted & 11.37\% & 1.57\% & -9.80\% \\
Drifted/Lost & 7.43\% & 0.95\% & -6.48\% \\
Semantic Drift & 0.104 & 0.006 & -94.70\% \\
Unsupported Detail Accretion & 0.311 & 0.021 & -93.20\% \\
Provenance Preservation & - & 91.45\% & - \\
\bottomrule
\end{tabularx}
\end{table}

\subsection{Heterogeneous Comparison}

Table~\ref{tab:heterogeneous-comparison} compares the heterogeneous baseline and provenance-aware results at stage five. At stage five, 95.73\% of provenance-aware outputs remain Equivalent to the original claim, compared with 31.33\% in the baseline. Expanded outputs decrease from 28.93\% to 0.67\%, contracted outputs from 6.73\% to 0.13\%, contradictions from 20.67\% to 2.07\% and drifted or lost outputs from 12.33\% to 1.40\%. The framework also reduces semantic drift by 96.4\% and unsupported detail accretion by 95.1\%, while 99.93\% of stage-five outputs pass all provenance checks. These results show that the implemented provenance-aware framework reduces claim changes and unsupported additions in the tested heterogeneous sequence. The increase in Equivalent outputs is larger than in the aggregated homogeneous results. However, this comparison does not establish that model heterogeneity itself produces greater benefits, because model identity, pipeline position and preceding transformations are intertwined in the fixed sequence. 

\begin{table}[H]
\caption{Heterogeneous Baseline and Provenance-Aware Comparison at Stage Five}
\label{tab:heterogeneous-comparison}
\centering
\tablestyle
\begin{tabularx}{\columnwidth}{@{}Lrrr@{}}
\toprule
Metrics & Baseline & \shortstack{Provenance-\\Aware} & Difference \\
\midrule
Equivalent & 31.33\% & 95.73\% & +64.40\% \\
Expanded & 28.93\% & 0.67\% & -28.26\% \\
Contracted & 6.73\% & 0.13\% & -6.60\% \\
Contradicted & 20.67\% & 2.07\% & -18.60\% \\
Drifted/Lost & 12.33\% & 1.40\% & -10.93\% \\
Semantic Drift & 0.194 & 0.007 & -96.40\% \\
Unsupported Detail Accretion & 0.590 & 0.029 & -95.10\% \\
Provenance Preservation & - & 99.93\% & - \\
\bottomrule
\end{tabularx}
\end{table}

\section{Conclusion}

As LLMs shift from single tools to multi-agent systems, the trustworthiness of their outputs depends on both what each model generates and what they share with each other. When claims pass between agents without citing the original source or support status, it becomes harder to distinguish inherited content from newly added information. The baseline results show that claim meaning can change and unsupported details can increase during repeated transmission. 

This research introduces a provenance-aware framework that keeps claim content separate from its support status, source, evidence, and transmission history. The baseline experiments show that unsupported claims change as they pass through multi-agent pipelines. These changes are larger in the tested heterogeneous sequence than in the aggregated homogeneous results.  In comparison, the provenance-aware framework helps agents preserve both the claim\textquoteright{}s actual meaning and its provenance across same model and mixed model workflows. The framework does not verify claims or block their transmission. Instead, it allows downstream agents and users to understand what supports a claim and how it reached them while reducing unsupported changes during processing.

The framework could support research-summary workflows in which one agent passes claims to a report-writing agent. Keeping each claim\textquoteright{}s support status and transmission history attached could help downstream agents or users identify unsupported statements for review before including them in a report. 

The findings show that the implemented provenance-aware framework preserves claim meaning and reduces measured unsupported additions relative to text-only transmission in the evaluated five-stage pipelines. Passing claims with their support status, source, evidence and transmission history also makes this information available for downstream review. These results apply to the homogeneous configurations and the fixed heterogeneous sequence tested here. Future work should examine alternative model orders, branching pipelines and feedback interactions. It should also evaluate the contribution of individual framework components and whether provenance remains intact across every stage of a complete chain.

\end{document}